\documentclass[%showpacs, showkeys,
12pt,
preprint,preprintnumbers,nofootinbib,
groupedaddress,superscriptaddress,amsmath,amssymb]{revtex4}
\usepackage{graphicx}% Include figure files
\usepackage{dcolumn}% Align table columns on decimal point
\usepackage{bm}% bold math
\usepackage{amssymb}
\usepackage{amsmath}
\usepackage{epsfig}    
\usepackage{color}
\usepackage{hhline}
\def\be{\begin{equation}}
\def\ee{\end{equation}}
\newcommand{\bea}{\begin{eqnarray}}
\newcommand{\eea}{\end{eqnarray}}
\newcommand{\nn}{\nonumber}

\numberwithin{equation}{section}

\begin{document}
{\begin{flushright}{ APCTP Pre2019 - 018}\end{flushright}}

%%%%%%%%%
\title{A radiative seesaw model in modular $A_4$ symmetry}

\author{Hiroshi Okada}
\email{hiroshi.okada@apctp.org}
\affiliation{Asia Pacific Center for Theoretical Physics (APCTP) - Headquarters San 31, Hyoja-dong,
Nam-gu, Pohang 790-784, Korea}

\author{Yuta Orikasa}
\email{Yuta.Orikasa@utef.cvut.cz}
\affiliation{Institute of Experimental and Applied Physics, 
Czech Technical University in Prague, 
Husova 240/5, 110 00 Prague 1, Czech Republic}

\date{\today}

\begin{abstract}
We study one-loop induced radiative seesaw model applying a modular $A_4$ flavor symmetry, 
in which the neutrino mass matrix is achieved by two different Yukawa couplings one of which also contributes to positive
value of muon anomalous magnetic moment as well as lepton flavor violations.   
Thanks to the specific mass matrix via $A_4$ symmetry and its modular weight, we
find several predictions for lepton sector through our numerical analysis.  
\end{abstract}
\maketitle
\newpage

\section{Introduction}
Radiative seesaw models provides rich phenomenologies at TeV scale such as dark matter (DM) candidate, lepton flavor violations (LFVs), muon anomalous magnetic moment, electroweak precision tests like Z-boson decays, and collider physics,
depending on models. Its representative scenario is known as Ma model~\cite{Ma:2006km}, which is the first model to correlate the neutrino sector and dark sector inside the neutrino mass loop.
%%%
Recently, modular flavor symmetries have been proposed~\cite{Feruglio:2017spp, deAdelhartToorop:2011re}
to provide more predictions to the quark and lepton sector, because any Yukawa couplings can have a representation of flavor groups as well as modular weight.~\footnote{One notices that Yukawa couplings have to have even number of modular weight only, especially the modular weight starts from 4 if the Yukawa coupling is assigned to be singlets under the flavor symmetry.}
Their typical groups are found in basis of  the $A_4$ modular group \cite{Feruglio:2017spp, Criado:2018thu, Kobayashi:2018scp, Okada:2018yrn, Nomura:2019jxj, Okada:2019uoy, deAnda:2018ecu, Novichkov:2018yse, Nomura:2019yft, Ding:2019xvi}, $S_3$ \cite{Kobayashi:2018vbk, Kobayashi:2018wkl, Kobayashi:2019rzp, Okada:2019xqk}, $S_4$ \cite{Penedo:2018nmg, Novichkov:2018ovf, Kobayashi:2019mna}, $A_5$ \cite{Novichkov:2018nkm, Ding:2019xna}, larger groups~\cite{Baur:2019kwi}, multiple modular symmetries~\cite{deMedeirosVarzielas:2019cyj}, and double covering of $A_4$~\cite{Liu:2019khw} in which  masses, mixings, and CP phases for quark and lepton are predicted.~\footnote{Several reviews are helpful to understand whole the ideas~\cite{Altarelli:2010gt, Ishimori:2010au, Ishimori:2012zz, Hernandez:2012ra, King:2013eh, King:2014nza, King:2017guk, Petcov:2017ggy}.}
Furthermore, thanks to the modular weight that is another degree of freedom originated from modular symmetry,
 this modular weight can be identified as a symmetry to stabilize DM candidate if DM is included in a model.
 Thus, radiative seesaw models with modular flavor symmetries are well motivated in view of neutrino predictions and DM origin.
%%%

In this paper, we apply a modular $A_4$ flavor symmetry to induce one-loop induced neutrino mass matrix with two different Yukawa couplings, running two singly-charged bosons and six singly-charged-fermions inside the neutrino loop.
Apart from the DM candidate this time, we try to derive positive muon anomalous magnetic moment from the same Yukawa coupling that contributes to the neutrino mass matrix as well as obtaining several predictions on the neutrino sector satisfying the LFVs through our numerical analysis.

This paper is organized as follows.
In Sec.~\ref{sec:realization},   we explain our model set up under modular $A_4$ symmetry, and formulate each of exotic mass matrix and mixing, LFVs, muon anomalous magnetic moment, the neutrino mass matrix and some relations to achieve the numerical analysis.
In Sec. III, we show numerical analysis and demonstrate several figures that provide predictions. 
Finally we conclude and discuss in Sec.~\ref{sec:conclusion}.

% \begin{widetext}
\begin{center} 
\begin{table}[tb]%[tbc]
%\begin{tiny}
\begin{tabular}{|c||c|c|c|c||c|c|c||}\hline\hline  
&\multicolumn{4}{c||}{ Fermions} & \multicolumn{3}{c||}{Bosons} \\\hline
  & ~$( L_{L_e}, L_{L_\mu}, L_{L_\tau})$~ & ~$(e_{R_e},e_{R_{\mu}},e_{R_{\tau}})$~& ~$L'_{1,2,3}$~ 
  & ~$e'_{1,2,3}$~ & ~$H$~  & ~$\eta$~  & ~$s^+$~
  \\\hline 
 %%%
 $SU(2)_L$ & $\bm{2}$  & $\bm{1}$  & $\bm{2}$ & $\bm{1}$   & $\bm{2}$  & $\bm{2}$ & $\bm{1}$    \\\hline 
 %%%
$U(1)_Y$ & -$\frac12$ & $-1$ & -$\frac12$  & $-1$& $\frac12$ & $\frac12$  & $1$        \\\hline
 %%%
 $A_4$ & $(1,1',1'')$ & $(1,1',1'')$ & $3$ & $3$ & $1$ & $1$ & $1$     \\\hline
 $-k$ & $0$ & $0$ & $-1$ & $-1$ & $0$ & $-1$ & $-3$    \\\hline
\end{tabular}
\caption{Field contents of fermions and bosons
and their charge assignments under $SU(2)_L\times U(1)_Y\times A_4$ in the lepton and boson sector, 
where $-k$ is the number of modular weight %, $a=1,2,3$, 
and the quark sector is the same as the SM.}
\label{tab:fields}
% \end{tiny}
\end{table}
\end{center}
%\end{widetext}

% \begin{widetext}
\begin{center} 
\begin{table}[tb]%[tbc]
%\begin{tiny}
\begin{tabular}{|c||c|c|c|c|c|c||}\hline\hline  
 &\multicolumn{3}{c||}{Couplings}  \\\hline
  & ~$Y^{(4)}_{\bf1}$~& ~$Y^{(2)}_{\bf3}$~ & ~$Y^{(4)}_{\bf3}$~ \\\hline 
 $A_4$ & ${\bf1}$ & ${\bf3}$& ${\bf3}$    \\\hline
 $-k$ & $4$ & $2$ & $4$     \\\hline
\end{tabular}
\caption{Modular weight assignments for Yukawa and Higgs couplings, the other couplings are all neutral under the modular {symmetry}. 
%, where $i=1,3$ denotes the component of $S_3$ triplet.
%Notice here that the number of modular weight for Higgs terms has to start at 4 because they are singlets under $A_4$ group.
}
\label{tab:couplings}
% \end{tiny}
\end{table}
\end{center}
%\end{widetext}

\section{ Model} 
\label{sec:realization}
In this section, we explain our model.
Here, we introduce three generation of exotic vector-like leptons $L'\equiv[N',E']^T$ and $e'$ under isospin doublet and singlet,
where both of them are assigned by $(3,-1)$ under ($A_4,-k$). Also we introduce an isospin doublet inert boson $\eta\equiv[\eta^+,\eta_0]^T$ and a singly-charged singlet boson $s^+$, where $\eta$ and $s^+$ are respectively assigned by $-1, -3$ under modular weight, but both of them are trivial singlet under $A_4$. The Standard Model (SM) leptons are assigned by $1,1',1''$ for $e,\mu,\tau$ under $A_4$ respectively, while they are neutral under modular weight. The SM Higgs is defined by $H$ and its VEV is denoted by $\langle H\rangle \equiv v_H/\sqrt2$, where $H$ is trivial singlet and neutral under $A_4$ and modular weight, respectively.
The {$A_4$ representation} and modular weight for fields
are given by Tab.~\ref{tab:fields}, while the ones of Yukawa couplings are respectively given by Tab.~\ref{tab:couplings}.
%%%
Under these symmetries, one writes renormalizable Lagrangian as follows:
\begin{align}
-{\cal L}_{Lepton} &=\sum_{\ell=e,\mu,\tau}y_\ell \bar L_{L_\ell} H e_{R_\ell}
+Y' (\bar L'_{L_e} H  e'_{R_e} +\bar L'_{L_\mu} H  e'_{R_\tau} +\bar L'_{L_\tau} H  e'_{R_\mu} )\nn\\
&+M' (\bar L'_{L_e} L'_{R_e} +\bar L'_{L_\mu} L'_{R_\tau} +\bar L'_{L_\tau} L'_{R_\mu}  )
+m' (\bar e'_{L_e} e'_{R_e} +\bar e'_{L_\mu} e'_{R_\tau} +\bar e'_{L_\tau} e'_{R_\mu}  )\nn\\
&+
\alpha_\eta(Y^{(2)}_{\bf3} \otimes \bar L_{L_e} \otimes e'_{R} ) \eta
+\beta_\eta(Y^{(2)}_{\bf3} \otimes \bar L_{L_\mu} \otimes e'_{R} ) \eta
+\gamma_\eta(Y^{(2)}_{\bf3} \otimes \bar L_{L_\tau} \otimes e'_{R} ) \eta\nn\\
&+
\alpha_s(Y^{(4)}_{\bf3} \otimes \bar L^C_{L_e} \otimes L'_{L} ) s^+
+\beta_s(Y^{(4)}_{\bf3} \otimes \bar L^C_{L_\mu} \otimes L'_{L} ) s^+
+\gamma_s(Y^{(4)}_{\bf3} \otimes \bar L^C_{L_\tau} \otimes L'_{L} ) s^+\nn\\
+ {\rm h.c.}, \label{eq:lag-lep}
\end{align}
where $Y',M',m'$ include the invariant modular factor $1/(-i\tau+i\bar\tau)$.

The  modular forms of weight 2, {$(y_{1},y_{2},y_{3})$},  transforming
as a triplet of $A_4$ is written in terms of Dedekind eta-function  $\eta(\tau)$ and its derivative \cite{Feruglio:2017spp}:
%%%%%%%%%%%%%%%%%%%%%%%
\begin{eqnarray} 
\label{eq:Y-A4}
y_{1}(\tau) &=& \frac{i}{2\pi}\left( \frac{\eta'(\tau/3)}{\eta(\tau/3)}  +\frac{\eta'((\tau +1)/3)}{\eta((\tau+1)/3)}  
+\frac{\eta'((\tau +2)/3)}{\eta((\tau+2)/3)} - \frac{27\eta'(3\tau)}{\eta(3\tau)}  \right), \nonumber \\
y_{2}(\tau) &=& \frac{-i}{\pi}\left( \frac{\eta'(\tau/3)}{\eta(\tau/3)}  +\omega^2\frac{\eta'((\tau +1)/3)}{\eta((\tau+1)/3)}  
+\omega \frac{\eta'((\tau +2)/3)}{\eta((\tau+2)/3)}  \right) , \label{eq:Yi} \\ 
y_{3}(\tau) &=& \frac{-i}{\pi}\left( \frac{\eta'(\tau/3)}{\eta(\tau/3)}  +\omega\frac{\eta'((\tau +1)/3)}{\eta((\tau+1)/3)}  
+\omega^2 \frac{\eta'((\tau +2)/3)}{\eta((\tau+2)/3)}  \right)\,.
\nonumber
\end{eqnarray}
%%%%%%%%%%%%%%%%%%%%%
%
% where 
The overall coefficient in Eq. (\ref{eq:Yi}) is 
one possible choice; it cannot be uniquely determined. 
%Thus we just impose the purtabative limit {$y_{1,2,3}\lesssim\sqrt{4\pi}$} in the numerical analysis.
%%%%%%%%%%%%%%%%%%%%%
Then, any couplings of higher weight are constructed by multiplication rules of $A_4$,
and one finds the following couplings:
%%%%%%%%%%%%%%%%%%%%%%%%%%
\begin{align}
&Y^{(4)}_{\bf1}=y^2_1+2y_2y_3,\quad
%Y^{(8)}_{\bf1}=(y^2_1+2y_2y_3)^2,\quad
%y^4_1-2y^2_1y_2y_3-2y_1y_3^3-2y_1y_2^3+2y_1^2y_2^2+3y_2^2y_3^2 ,\nn\\
Y^{(4)}_{\bf3}\equiv
\left[\begin{array}{c}
y'_1 \\ 
y'_2 \\ 
y'_3 \\ 
\end{array}\right]
=
\left[\begin{array}{c}
y^2_1-y_2y_3 \\ 
y^2_3-y_1y_2 \\ 
y^2_2-y_1y_3 \\ 
\end{array}\right].
\end{align}
%%%%%%%%%%%%%%%%%%%%%%%
%%%%%%%%%%%%%%%%%%%

\if0
Higgs potential is found in a 
\begin{align}
{\cal V} &= -\mu_H^2 |H|^2 +\mu^2_\eta |Y^{(4)}_{\bf1}||\eta|^2\\
&+ \frac14 \lambda_H|H|^4+ \frac14\lambda_\eta |Y^{(8)}_{\bf1}| |\eta|^4
+\lambda_{H\eta} |Y^{(4)}_{\bf1}||H|^2|\eta|^2+\lambda_{H\eta}' |Y^{(4)}_{\bf1}| |H^\dag\eta|^2
+\frac14\lambda_{H\eta}'' [Y^{(4)}_{\bf1}(H^\dag\eta)^2+ {\rm h.c.}]\nn,
 \label{eq:pot}
\end{align}
which can be the same as the original potential of Ma model without loss of generality, because of additional free parameters.
The point is that one does not have a term $H^\dag\eta$ due to absence of $S_3$ singlet with modular weight $2$ that arises from the feature of modular symmetry.
\fi

\subsection{ Singly-charged exotic fermion mass matrix}
After the electroweak spontaneous symmetry breaking,  singly-charged exotic fermion mass matrix is found in basis of $(e'^-,E'^-)_R$ as
\begin{align}
{\cal M}_E &=%\frac1{2}
\left[\begin{array}{cc}
m' P_{23} &m_{e'L'} P_{23} \\ 
m_{e'L'}^\dag P_{23}  & M' P_{23}   \\ 
\end{array}\right],\quad
%%%
P_{23}=
\left[\begin{array}{ccc}
1 & 0 & 0 \\ 
0 & 0 & 1 \\ 
0 & 1 & 0 \\ 
\end{array}\right],
\label{eq:mn}
\end{align}
where $m_{e'L'}\equiv v_H Y'/\sqrt2$.
Then, ${\cal M}_E $ is diagonalized by a unitary mixing matrix; $D_E\equiv V^* {\cal M}_E V^T$, where two set of three degenerate mass eigenstates are given by this mass matrix,
and the mass eigenstates are related to the flavor eigenstates $\psi$ as follows: $(e'^-,E'^-)^T _{L,R}\equiv V^T \psi_{L,R}^-$.

\subsection{ Neutral exotic fermion mass matrix}
Similar to the singly-charged exotic fermion mass matrix, its form is found as
\begin{align}
{\cal M}_N &=%\frac1{2}
 M' 
\left[\begin{array}{ccc}
1 & 0 & 0 \\ 
0 & 0 & 1 \\ 
0 & 1 & 0 \\ 
\end{array}\right].
\label{eq:mn}
\end{align}
Then, ${\cal M}_N $ is diagonalized by the unitary mixing matrix; $D_N\equiv V_N^* {\cal M}_N V^T_N$, where three degenerate mass eigenstates are given by this mass matrix,
and the mass eigenstates are related to the flavor eigenstates $N'$ as follows: $N'^T _{L,R}\equiv V^T_{N} \psi_{L,R}^0$.

\subsection{Singly-charged bosons}
The singly-charged bosons mix each other through the term of $Y^{(4)}_1(\eta^T\cdot H) s^+$, where $\sigma_2$ is the second Pauli matrix. Here, we define as follows:
\begin{align}
s^\pm=c_\alpha H_1^\pm + s_\alpha H_2^\pm,\quad \eta^\pm =-s_\alpha H_1^\pm + c_\alpha H_2^\pm,
\end{align}
where $s_\alpha(c_\alpha)$ is the short-hand symbol of $\sin \alpha(\cos \alpha)$, and $\cdot\equiv i\sigma_2$.

%%%%%%%%%%%%
\subsection{Neutrino mass matrix and lepton flavor violations}
Neutrino mass matrix and LFVs are originated from the terms of $\bar L_L \eta e'_R$ and $\bar L^C_L L'_L s^+$,
and their explicit forms are given by 
\begin{align}
{\cal L}&= \sum_{i=1}^3 \sum_{a=1}^3 \sum_{b=1}^6 \left[\bar \nu_{L_i} (y_\eta)_{i, a}  (V^T)_{a, b} \psi_{R_b}^- (-s_\alpha H_1^+ + c_\alpha H_2^+)
+
 \bar \nu_{L_i}^C (y_s)_{i, a} (V^T)_{a+3, b} \psi_{L_b }^- (c_\alpha H_1^+ + s_\alpha H_2^+)\right] \nn\\
 %%%
&+ \sum_{i=1}^3 \sum_{a=1}^3 \sum_{b=1}^6 \bar \ell_{L_i} (y_\eta)_{i, a}  (V^T)_{a, b} \psi_{R_b}^- \eta_0
+
\sum_{i=1}^3 \sum_{j=1}^3 \bar \ell_{L_i}^C (y_s V^T_N)_{ij} \psi^0_{L_j} (c_\alpha H_1^+ + s_\alpha H_2^+)
 +{\rm h.c.},
\end{align}
where 
%$V'^T_{R}\equiv (V^T_{R})_{i,a}$ and $V'^T_{L} \equiv (V^T_{L})_{i+3,a}$ with $i=1-3$ and $a=1-6$, and 
%
\begin{align}
y_\eta &=%\frac1{2}
\left[\begin{array}{ccc}
\alpha_\eta & 0 & 0 \\ 
0 & \beta_\eta & 0 \\ 
0 & 0 & \gamma_\eta \\ 
\end{array}\right]
%%%
\left[\begin{array}{ccc}
y_1 &y_3 &y_2 \\ 
y_2 &y_1 &y_3 \\ 
y_3 &y_2 &y_1 \\ 
\end{array}\right],\
%%% %%%
y_S =%\frac1{2}
\left[\begin{array}{ccc}
\alpha_s & 0 & 0 \\ 
0 & \beta_s & 0 \\ 
0 & 0 & \gamma_s \\ 
\end{array}\right]
%%%
\left[\begin{array}{ccc}
1 & 0 & 0 \\ 
0 & 0 & 1 \\ 
0 & 1 & 0 \\ 
\end{array}\right]
%%%
\left[\begin{array}{ccc}
y'_1 &y'_3 &y'_2 \\ 
y'_2 &y'_1 &y'_3 \\ 
y'_3 &y'_2 &y'_1 \\ 
\end{array}\right]  .
\label{eq:mn}
\end{align}

{\it Lepton flavor violations} also arises from $y_D$ as~\cite{Baek:2016kud, Lindner:2016bgg}
\begin{align}
&{\rm BR}(\ell_i\to\ell_j\gamma)\approx\frac{48\pi^3\alpha_{em}C_{ij}}{G_F^2 (4\pi)^4}|{\cal M}_{i, j}|^2,\nn\\
&
{\cal M}_{i, j}  \approx
- \sum_{a=1}^3 \sum_{b=1}^6 \sum_{c=1}^3 (y_{\eta})_{j, a} (V^T)_{a, b} (V^\ast )_{b, c} (y_{\eta}^\dag)_{c, i}  F(\psi^-_b,\eta_0)
\nn\\
&+
\sum_{a=1}^3 \sum_{b=1}^6 \sum_{c=1}^3 (y_s)_{i, a}  (V^T_N)_{a+3, b} (V^\ast_N)_{b, c+3} (y_s^\dag)_{c, j}  \left[c^2_\alpha F(\psi_b^0 , H_1^-) + s^2_\alpha F(\psi_b^0 , H_2^-)\right],\\
%%%
%%%
&F(a,b)\approx\frac{2 m^6_a+3m^4_am^2_b-6m^2_am^4_b+m^6_b+12m^4_am^2_b\ln\left(\frac{m_b}{m_a}\right)}{12(m^2_a-m^2_b)^4},
\end{align}
where $C_{21}=1$, $C_{31}=0.1784$, $C_{32}=0.1736$, $\alpha_{em}(m_Z)=1/128.9$, and $G_F=1.166\times10^{-5}$ GeV$^{-2}$.
The experimental upper bounds are given by~\cite{TheMEG:2016wtm, Aubert:2009ag,Renga:2018fpd}
\begin{align}
{\rm BR}(\mu\to e\gamma)\lesssim 4.2\times10^{-13},\quad 
{\rm BR}(\tau\to e\gamma)\lesssim 3.3\times10^{-8},\quad
{\rm BR}(\tau\to\mu\gamma)\lesssim 4.4\times10^{-8},\label{eq:lfvs-cond}
\end{align}
which will be imposed in our numerical calculation.
 Muon g-2 is also given by
 \begin{align}
 \Delta a_\mu\approx -2\frac{m_\mu^2}{(4\pi)^2} {\cal M}_{2, 2}.
 \end{align}
 It implies that $y_\eta$ term provides positive contribution, while $y_s$ does negative contribution,
 since the experimental result suggests positive anomaly, $y_\eta>>y_s$ is expected. 

\if0
 %%%%%%%%%%%%%%%%%%%
\begin{figure}[tb]\begin{center}
\includegraphics[width=100mm]{diagram1.eps}
\caption{One loop diagram generating neutrino mass.}   
\label{fig:diagram}\end{center}\end{figure}
%%%%%%%%%%%%%%%%%%%
\fi

Neutrino mass matrix  is given at one-loop level, and its form is found as
\begin{align}
&m_{\nu_{ij}}\approx s_\alpha c_\alpha \sum_{a=1-6}
\frac{Y_{\eta_{ia}} {D_E}_{a} Y^\dag_{S_{a j}} + Y_{S_{ia}}^* {D_E}_{a} Y^T_{\eta_{a j}}}{(4\pi)^2}
%%%
\left(\frac{m_1^2}{m_1^2-{ D_E^2}_{a}}\ln\left[\frac{m_1^2}{{ D_E^2}_{a}}\right]
-
\frac{m_2^2}{m_2^2-{D_E}_{a}}\ln\left[\frac{m_2^2}{{ D_E^2}_{a}}\right]
\right),
\end{align}
where $(Y_\eta)_{i, j} \equiv \sum_{a=1}^3 (y_\eta)_{i, a} (V^T)_{a, j}$, $(Y_S)_{i, j} \equiv \sum_{a=1}^3 (y_s)_{i, a} (V^T)_{a+3, j}$, {$m_{1,2}$} is the mass of  $H^\pm_{1,2}$.
Then the neutrino mass matrix is diagonalized by an unitary matrix $U_{\nu}$ as $U_{\nu}m_\nu U^T_{\nu}=$diag($m_{\nu_1},m_{\nu_2},m_{\nu_3}$)$\equiv D_\nu$, where Tr$[D_{\nu}] \lesssim$ 0.12 eV is given by the recent cosmological data~\cite{Aghanim:2018eyx}.
Then, one finds $U_{PMNS}=V^\dag_{eL} U_\nu$.
Each of mixing is given in terms of the component of $U_{MNS}$ as follows:
\begin{align}
\sin^2\theta_{13}=|(U_{PMNS})_{13}|^2,\quad 
\sin^2\theta_{23}=\frac{|(U_{PMNS})_{23}|^2}{1-|(U_{PMNS})_{13}|^2},\quad 
\sin^2\theta_{12}=\frac{|(U_{PMNS})_{12}|^2}{1-|(U_{PMNS})_{13}|^2}.
\end{align}
Also, the effective mass for the neutrinoless double beta decay is given by
\begin{align}
m_{ee}=|D_{\nu_1} \cos^2\theta_{12} \cos^2\theta_{13}+D_{\nu_2} \sin^2\theta_{12} \cos^2\theta_{13}e^{i\alpha_{21}}
+D_{\nu_3} \sin^2\theta_{13}e^{i(\alpha_{31}-2\delta_{CP})}|,
\end{align}
where its observed value could be measured by KamLAND-Zen in future~\cite{KamLAND-Zen:2016pfg}.

To achieve numerical analysis, we derive several relations.
First of all, we define the normalized neutrino mass matrix as follows:
%\begin{align}
%&\tilde m_{\nu_{ij}}\equiv \frac{m_{\nu_{ij}}}{k_6}\approx \sum_{a=1-6}
%\frac{Y_{\eta_{ia}} F_{a} Y^\dag_{S_{a j}} + Y_{S_{ia}}^* F_{a}  Y^T_{\eta_{a j}}}{(4\pi)^2},\quad
%k_a\equiv s_\alpha c_\alpha \tilde k_a,\nn\\
%%
%\tilde k_a&\equiv D_{E_a} 
%\left(\frac{m_1^2}{m_1^2-{ D_E^2}_{a}}\ln\left[\frac{m_1^2}{{ D_E^2}_{a}}\right]
%-
%\frac{m_2^2}{m_2^2-{D_E}_{a}}\ln\left[\frac{m_2^2}{{ D_E^2}_{a}}\right]
%\right),\
%%
%F_a \equiv \frac{k_a}{k_6}.\label{eq:norm-nu}
%\end{align}
%Then the normalized neutrino mass eigenvalues are given in terms of neutrino mass eigenvalues; diag$(\tilde m_{\nu_1}^2,\tilde m_{\nu_2}^2,\tilde m_{\nu_3}^2)={\rm diag}(m_{\nu_1}^2,m_{\nu_2}^2,m_{\nu_3}^2)/k_6^2$. 
% It is found that $k_6^2$ is given by
%\begin{align}
%k_6^2=\frac{\Delta m^2_{\rm atm}}{\tilde m_{\nu_3}^2 - \tilde m_{\nu_1}^2},\label{eq:k}
%\end{align}
%where normal hierarchy is assumed and $\Delta m^2_{\rm atm}$ is the atmospheric neutrino mass difference square.
%Comparing Eq.(\ref{eq:norm-nu}) and Eq.(\ref{eq:k}), we find $s^2_\alpha$ is rewritten by the other parameters as follows:
%\begin{align}
%\tilde k_6^2 (s^2_\alpha)^2 - \tilde k_6^2 s^2_\alpha + k_6^2=0 ,\quad
%s^2_\alpha= \frac{\tilde k_6^2\pm\sqrt{\tilde k_6^4 - 4\tilde k_6^2 k_6^2}}{2 \tilde k_6^2},
% \label{eq:sa}
%\end{align}
%where $0\le s^2_\alpha\le 1$ should be satisfied.
% The solar neutrino mass difference square is also found as 
% \begin{align}
%\Delta m^2_{\rm sol}=\Delta m^2_{\rm atm} \frac{\tilde m_{\nu_2}^2 - \tilde m_{\nu_1}^2}{\tilde m_{\nu_3}^2 - \tilde m_{\nu_1}^2},\label{eq:k}
%\end{align}
\begin{align}
&\tilde m_{\nu_{ij}}\equiv \frac{m_{\nu_{ij}}}{s_\alpha c_\alpha}\approx \sum_{a=1-6}
\frac{Y_{\eta_{ia}} F_{a} Y^\dag_{S_{a j}} + Y_{S_{ia}}^* F_{a}  Y^T_{\eta_{a j}}}{(4\pi)^2},\nn\\
F_a &\equiv D_{E_a} 
\left(\frac{m_1^2}{m_1^2-{ D_E^2}_{a}}\ln\left[\frac{m_1^2}{{ D_E^2}_{a}}\right]
-
\frac{m_2^2}{m_2^2-{D_E}_{a}}\ln\left[\frac{m_2^2}{{ D_E^2}_{a}}\right]
\right). \label{eq:norm-nu}
\end{align}
Then the normalized neutrino mass eigenvalues are given in terms of neutrino mass eigenvalues; diag$(\tilde m_{\nu_1}^2,\tilde m_{\nu_2}^2,\tilde m_{\nu_3}^2)={\rm diag}(m_{\nu_1}^2,m_{\nu_2}^2,m_{\nu_3}^2)/(s_\alpha c_\alpha)^2$. 
 It is found that $s_\alpha$ is given by
\begin{align}
s_\alpha^2 (1 - s_\alpha^2)=\frac{\Delta m^2_{\rm atm}}{\tilde m_{\nu_3}^2 - \tilde m_{\nu_1}^2},\label{eq:k}
\end{align}
where normal hierarchy is assumed and $\Delta m^2_{\rm atm}$ is the atmospheric neutrino mass difference square.
The solar neutrino mass difference square is also found as 
 \begin{align}
\Delta m^2_{\rm sol}=\Delta m^2_{\rm atm} \frac{\tilde m_{\nu_2}^2 - \tilde m_{\nu_1}^2}{\tilde m_{\nu_3}^2 - \tilde m_{\nu_1}^2},\label{eq:k}
\end{align}
In numerical analysis, this value should be within the experimental result, while $\Delta m^2_{\rm atm}$ is expected to be input parameter.

\section{Numerical analysis}
Here, we show numerical analysis to satisfy all of the constraints that we discussed above,
where we assume the neutrino mass ordering is normal hierarchy and {$m_{\eta^\pm} \approx m_2$} to avoid the oblique parameters simply.
%In this case, the mass of DM is uniquely fixed by the observed relic density which suggests it is within $534\pm8.5$ GeV~\cite{Hambye:2009pw}, if the Yukawa coupling is not so large. In fact, tiny Yukawa couplings are requested by satisfying the data. Thus, we just work on the mass of $\eta$ at this narrow range.~\footnote{We have checked there exist allowed region to satisfy the neutrino oscillation data and LFVs in the case of fermionic DM case; $N_{R}$, and its mass is 18$-$19 GeV. However we cannot explain the observed relic density, since the Yukawa couplings are too tiny.}

Then, we provide the experimentally allowed ranges for neutrino mixings and mass difference squares at 3$\sigma$ range~\cite{Esteban:2018azc} as follows:
\begin{align}
&\Delta m^2_{\rm atm}=[2.431-2.622]\times 10^{-3}\ {\rm eV}^2,\
\Delta m^2_{\rm sol}=[6.79-8.01]\times 10^{-5}\ {\rm eV}^2,\label{eq:exps} \\
&\sin^2\theta_{13}=[0.02044-0.02437],\ 
\sin^2\theta_{23}=[0.428-0.624],\ 
\sin^2\theta_{12}=[0.275-0.350].\nn
\end{align}
{The range of absolute value of the complex dimensionless parameters $\alpha_\eta,\beta_\eta, \gamma_\eta$ are taken to be [0.1-1], while $\alpha_s,\beta_s, \gamma_s$ are taken to be $[10^{-10},10^{-5}]$ with real parameter.}
\footnote{The large hierarchies between $\alpha_{\eta},\beta_\eta,\gamma_\eta$ and $\alpha_s,\beta_s, \gamma_s$ are expected to induce positive muon $g-2$ as discussed in the part of $\Delta a_\mu$.}
The mass parameters $m_{1,2},m',m_{e'L'},M'$ are $[0.1,10]$ TeV.

%%%%%%%%%%%%%%%%%%%
\begin{figure}[tb]\begin{center}
\includegraphics[width=80mm]{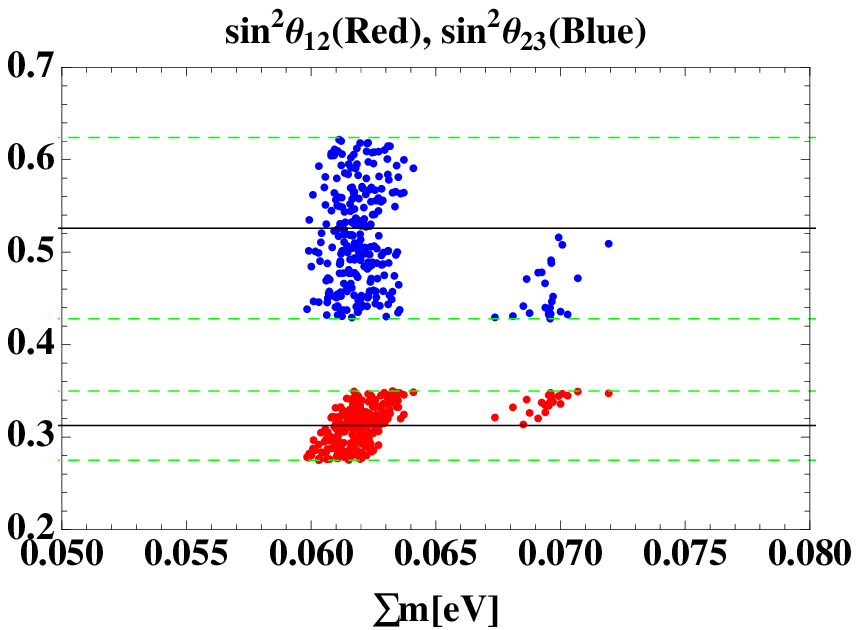}
\caption{The sum of neutrino masses $\sum m(\equiv$ Tr$[ D_\nu]$) versus $\sin^2\theta_{12}$(red color) and  $\sin^2\theta_{23}$(blue color). Here, the horizontal black solid lines are the best fit values, the green dotted lines show 3$\sigma$ range. 
%, and the vertical black line shows upper bound on the cosmological data as shown in the neutrino section.
}   
\label{fig:1}\end{center}\end{figure}
%%%%%%%%%%%%%%%%%%%

Fig.~\ref{fig:1} shows the sum of neutrino masses $\sum m(\equiv$ Tr$[D_\nu])$ versus $\sin^2\theta_{12}$(red color) and  $\sin^2\theta_{23}$(blue color).
Here, the horizontal black solid lines are the best fit values, the green dotted lines show 3$\sigma$ range. 
%, and the vertical black line shows upper bound on the cosmological data as shown in the neutrino section.
Three of the mixing angles satisfy whole the allowed region at 3 $\sigma$ interval, but $ \sin^2 \theta_{23}$
tends to be in favor of the lower range [0.428-0.522], while $\sin^2 \theta_{12}$
tends to be in favor of the upper range [0.312-0.350].
The sum of neutrino masses are within the range of [0.060-0.064, 0.067-0.072], which is below the cosmological bound 0.12.
Here, $\sin^2 \theta_{13}$ runs whole the range of the allowed region in Eq.~(\ref{eq:exps}) without any favored regions.

%%%%%%%%%%%%%%%%%%%
\begin{figure}[tb]\begin{center}
\includegraphics[width=80mm]{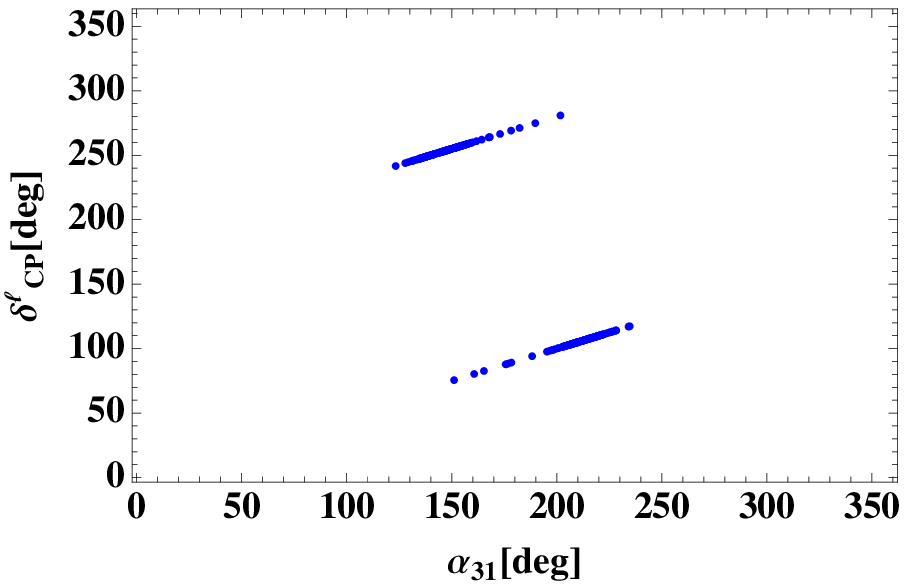}
\caption{Phases of $\delta^\ell_{CP}$ in terms of $\alpha_{31}$, where $\alpha_{21}$ is found to be zero.}   
\label{fig:2}\end{center}\end{figure}
%%%%%%%%%%%%%%%%%%%

Fig.~\ref{fig:2} shows phases of $\delta^\ell_{CP}$ in terms of $\alpha_{31}$.
This figure implies that the Dirac CP phase is favored in the range of [70-120, 240-280][deg], and $\alpha_{31}$ is favored in the range of [120-280][deg], where $\alpha_{21}$ is found to be zero.
Since the Dirac CP phase of $3\pi/2$(=270 [deg]) is favored by T2K experiment, our model could be well predicted and tested further.

%%%%%%%%%%%%%%%%%%%
\begin{figure}[tb]\begin{center}
\includegraphics[width=80mm]{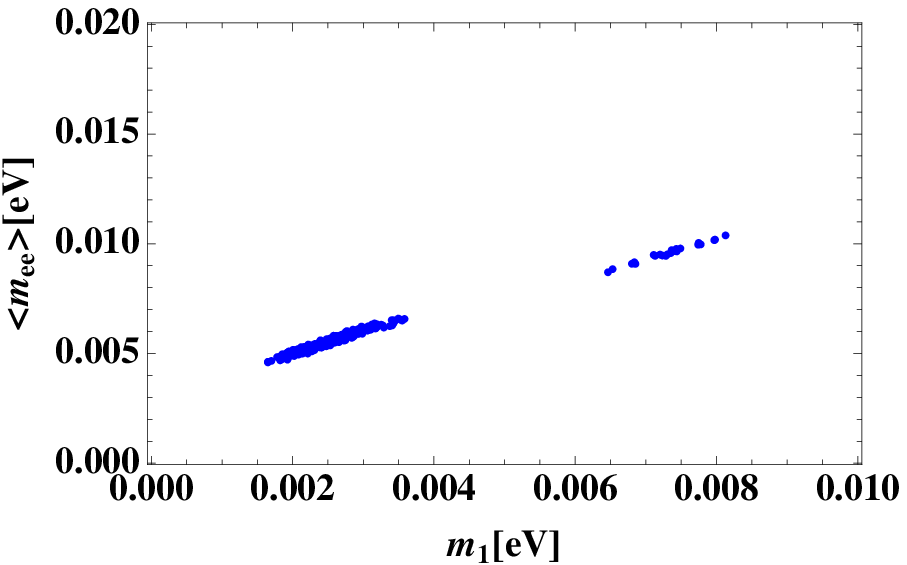}
\caption{The lightest neutrino mass versus {the effective mass for the} neutrinoless double beta decay.}   
\label{fig:3}\end{center}\end{figure}
%%%%%%%%%%%%%%%%%%%

Fig.~\ref{fig:3} demonstrates the lightest neutrino mass versus {the effective mass for the} neutrinoless double beta decay.
It suggests that $m_1$ is in the range of [0.0015-0.035,0.0065-0.0085] eV while  $m_1$ is in the range of
 [0.0045-0.0065, 0.008-0.010] eV.
The other remarks are in order:
 \begin{enumerate}
\item
The typical region of modulus $\tau$ is found in narrow space as 
\[ 1.85\ \lesssim\ {\rm Re} [\tau]\lesssim\ 1.95\ {\rm and}\  1.65\ \lesssim\ {\rm Im}[\tau]\lesssim\ 1.85.\]
\item 
The upper bound of our LFVs are  as follows:
\[{\rm BR}(\mu\to e\gamma)\lesssim4.2\times10^{-13}, \quad {\rm BR}(\tau\to e\gamma)\lesssim1.2\times10^{-13},\quad {\rm BR}(\tau\to \mu\gamma)\lesssim4\times10^{-13}.\]
These imply that the future experiment of $\mu\to e\gamma$ is promising to test our model.
\item The muon anomalous magnetic moment is positively obtained by the scale cannot be so large enough to reach the experimentally expected value by five order magnitude. Our upper bound of $\Delta a_\mu$ is found to be
   \[\Delta a_\mu\lesssim 6\times 10^{-14}.\]
 \end{enumerate}

\section{Conclusion and discussion}
\label{sec:conclusion}
We have constructed a predictive lepton model with modular $A_4$ symmetry in framework of one-loop induced radiative seesaw model, running singly-charged bosons and singly-charged-fermions. We have the term that can contribute to the muon anomalous magnetic moment with positive value as well as the one with negative value, and these terms also provide us the structure of neutrino mass matrix. The each structure of mass matrix for charged-leptons and neutrinos is uniquely determined by the $A_4$ symmetry,
and inert property is assured by modular weight. Especially, the charged-lepton mass matrix can always be diagonal in the stage of flavor eigenstate thanks to the $A_4$ symmetry, once we assign each of field  $(e,\mu,\tau)$ to $(1,1',1'')$.
Due to the $A_4$ symmetry and their assignments, we have found the exotic fermion mass structures are very simple and two sets of three degenerate charged-fermion mass eigenvalues, and three degenerate neutral heavy fermion mass eigenvalues.
In numerical analysis, we have found only $\mu\to e\gamma$ reached the current upper bound and future experiment 
will provides more restriction for our model. Even though we have gotten the positive muon $g-2$, but the scale is very small compared to the experimentally expected result. Our maximum value is around $6\times 10^{-14}$ that is as tiny as the experimental value by five order magnitude.     
 %
%Thanks to the nonzero modular weight, the DM stability is naturally assured by this new quantum number, and DM is correlated with neutrinos in a specific manner, where their interactions are determined by the $S_3$ symmetry that is known as the minimal group in non-Abelian discrete flavor symmetries. 
In our numerical analyses, we have found several remarkable predictions on neutrino sector as follows: 
 \begin{enumerate}
\item
 Three of the mixing angles satisfy whole the allowed region at 3 $\sigma$ interval, but $ \sin^2 \theta_{23}$
tends to be in favor of the lower range [0.428-0.522], while $\sin^2 \theta_{12}$
tends to be in favor of the upper range [0.312-0.350].
The sum of neutrino masses are within the range of [0.060-0.064, 0.067-0.072], which is below the cosmological bound 0.12.
\item The Dirac CP phase is favored in the range of [70-120, 240-280][deg], and $\alpha_{31}$ is favored in the range of [120-280][deg], where $\alpha_{21}$ is found to be zero.
Since the Dirac CP phase of $3\pi/2$(=270 [deg]) is favored by T2K experiment, our model could be well predicted and tested further.
 \end{enumerate}
These predictions will be tested in the near future.

%\newpage
%%%%%%%%%%%%%%%%%%%%%%%%%%%%%%%%%%%
\section*{Acknowledgments}
\vspace{0.5cm}
{\it
This research was supported by an appointment to the JRG Program at the APCTP through the Science and Technology Promotion Fund and Lottery Fund of the Korean Government. This was also supported by the Korean Local Governments - Gyeongsangbuk-do Province and Pohang City (H.O.). 
H. O. is sincerely grateful for the KIAS member, and log cabin at POSTECH to provide nice space to come up with this project. 
Y. O. was supported from European Regional Development Fund-Project Engineering Applications of Microworld
Physics (No.CZ.02.1.01/0.0/0.0/16\_019/0000766)}
%%%%%%%%%%%%%%%%%%%%%%%%%%%%%%%%%%%
%%%%%%%%%%%%%%%%%%%%%%%%%%%%%%%%%%%

%\section*{Appendix}...

\end{document}